\documentclass[aps,twocolumn,showpacs,groupedaddress,nofootinbib]{revtex4-1}
\usepackage{epsfig}
\usepackage{bm}
\usepackage{amsmath,amssymb,times}

\newcommand{\ds}{\displaystyle}

\newcommand{\be}{\begin{equation}}
\newcommand{\ee}{\end{equation}}
\newcommand{\ba}{\begin{eqnarray}}
\newcommand{\ea}{\end{eqnarray}}
\newcommand{\etal}{\mbox{\it et al.}}

\begin{document}

\preprint{02-02}

\title{Neutral pion electroproduction in $p(e,e'\pi^0)p$ above $\sqrt{s}>2$~GeV} 
\author{Murat M. Kaskulov}
\email{murat.kaskulov@theo.physik.uni-giessen.de}
\affiliation{Institut f\"ur Theoretische Physik, Universit\"at Giessen,
             D-35392 Giessen, Germany}
\date{\today}

\begin{abstract}
Electroproduction of neutral pions in exclusive reaction $p(e,e'\pi^0)p$ 
is studied above the resonance region, $\sqrt{s}>2$~GeV. The reaction amplitude
is described by exchanges of vector $\omega(782)$, $\rho(770)$ and
axial-vector $h_1(1170)$  and $b_1(1235)$ Regge trajectories. The residual 
effect of $s$- and $u$-channel nucleon resonances is taken into account using a dual
connection between the exclusive form factors and inclusive deep inelastic 
structure functions. In photoproduction  at forward angles the exchange of Regge trajectories 
dominates and the dip region is filled by the resonances. 
In electroproduction  the excitation of nucleon resonances explains the high 
$Q^2$ data from JLAB.  The results for the beam spin azimuthal asymmetry measured 
at CLAS/JLAB are presented. Model calculations in the deep inelastic region at 
HERMES/DESY are given.
\end{abstract}
\maketitle


\section{Introduction}
Electroproduction of mesons in the deep inelastic scattering (DIS),
that is $\sqrt{s}>2$~GeV and $Q^2>1$~GeV$^2$, 
is a modern tool which permits to study the structure of the nucleon on the partonic
level. Exclusive channels in DIS are of particular importance. 
In this kind of hard exclusive processes one may learn about the off-forward 
parton distributions  that parameterize an intrinsic
nonperturbative pattern of the nucleon, see Ref.~\cite{Weiss:2009ar} and references
therein. Much work have been done to 
understand the production of pions in exclusive kinematics.
For instance, in QCD at large values of $(\sqrt{s},Q^2)$
and finite value of Bjorken $x_{\rm B}$  the description of $N(e,e'\pi)N'$ 
relies on the dominance of the longitudinal cross 
section $\sigma_{\rm L}$~\cite{Collins:1996fb}.  The transverse cross section
$\sigma_{\rm T}$ is
predicted to be  suppressed by power of $\sim 1/Q^2$. 
However, being a leading twist prediction the kinematic domain where this
power suppression starts to dominate is not
yet known for exclusive $\pi$ production. 

A somewhat different concept is used in the Regge pole models which rely on 
effective degrees of freedom. Here the forward $(\gamma^*,\pi)$ 
production mechanism is peripheral, that is a sum of all possible
$t$-channel meson-exchange processes. Although both partonic and Regge 
descriptions are presumably dual the ongoing and planned experiments 
have a potential to discriminate between different models. The  related 
studies have been carried out at 
JLAB~\cite{Horn:2007ug,Blok:2008jy,Collaboration:2010kna} and at DESY~\cite{:2007an}.   
A dedicated program on exclusive production of pions is planned in the future 
at the JLAB upgrade~\cite{Horn:2009dn}.

A closely related phenomenon is the color
transparency (CT) effect. It becomes effective in exclusive electroproduction of mesons
off nuclei, see Ref.~\cite{:2007gqa} for a possible observation and
Refs.~\cite{Larson:2006ge,Cosyn:2007er,Kaskulov:2008ej} for further interpretations of the
CT signal in the reaction $A(e,e'\pi)$. Presently one believes that at high
values of $Q^2$  the exclusive pions are produced in point like
configurations which may interact  in the nuclear medium only weakly. 
However, the fate of pions inside the nucleus depends on the initial longitudinal
and/or transverse production mechanisms~\cite{Kaskulov:2008ej}.
Therefore, our
understanding of the
$\pi$ production off nucleons is mandatory for a proper interpretation of the CT
signal observed in electroproduction off nuclei.

On the experimental side, it would be interesting to see an onset of 
$\sigma_{\rm L}/\sigma_{\rm T} \propto Q^2$ scaling already at presently
avaliable energies. However, the high $Q^2$ data from JLAB~\cite{Horn:2006tm} 
and single spin asymmetries measured in the true DIS region at
HERMES~\cite{:2009ua} demonstrate the presence of nonvanishing transverse components 
in $p(\gamma^*,\pi^+)n$. Moreover, the high $Q^2$ region at JLAB~\cite{Horn:2006tm,:2009ub}, 
DESY~\cite{Desy,Ackermann:1977rp,Brauel:1979zk}, 
Cornell~\cite{Cornell_1,Cornell_2,Cornell_3} and CEA~\cite{{CEA}}   
is even dominated by the conversion of transverse photons.  
For instance, the $Q^2$ dependence of $\sigma_{\rm L}$ and $\sigma_{\rm T}$
in the $\pi^{+}$ electroproduction above $\sqrt{s}>2$~GeV  has been studied
in~\cite{Horn:2007ug}.  In the charged pion case  the longitudinal cross 
section $\sigma_{\rm L}$ at forward angles is well described by the 
quasi-elastic  $\pi$ knockout mechanism~\cite{Sullivan:1970yq,Neudatchin:2004pu}. 
It is driven by the pion charge form factor both at 
JLAB~\cite{Horn:2006tm,Tadevosyan:2007yd,Huber:2008id} and 
HERMES~\cite{Kaskulov:2009gp}. On the contrary, the $(\sqrt{s},Q^2)$ 
behavior of $\sigma_{\rm T}$ remains to be puzzling. The data demonstrate 
that $\sigma_{\rm T}$ is large and tends to increase relative to
$\sigma_{\rm L}$ as a function of $Q^2$. Interestingly, the $(\sqrt{s},Q^2)$
dependence of exclusive $\sigma_{\rm T}$ exhibits features of the 
semi-inclusive  $p(e,e'\pi)X$ reaction in DIS in the limit $z\to
1$~\cite{Kaskulov:2008xc}. This kind of an exclusive-inclusive 
connection~\cite{Bjorken:1973gc} has been also observed in exclusive 
$(\gamma^*,\rho^0)$ production~\cite{Gallmeister:2010wn}.

Theoretically, hadronic models based on the meson-exchange 
scenario alone largely underestimate the measured $\sigma_{\rm T}$ 
in charged pion electroproduction, see Ref.~\cite{Blok:2008jy} for 
further discussions and references therein. In the partonic models 
the higher twist transversity distributions might be relevant to 
understand the $(\gamma_{\rm T},\pi)$ reaction~\cite{Ahmad:2008hp,Goloskokov:2009ia}. 
Several phenomenological models already attempted to describe 
the $\sigma_{\rm T}$  component~\cite{Faessler:2007bc,Kaskulov:2008xc,KM}. 
For instance, the description of charged pion production proposed in Ref.~\cite{KM} relies on
the residual contribution of the nucleon resonances. This approach will be followed in
this work. It is supposed that the excitations of nucleon resonances dominate
in electroproduction. The resonances are dual to the direct partonic
interactions due to the Bloom-Gilman duality
connection~\cite{Bloom:1971ye,Bloom:1970xb}. 
In~\cite{KM} the resonances supplement the Regge based 
exchanges of single mesonic
trajectories. Therefore, one distinguishes the
peripheral $t$-channel meson-exchange processes and the $s(u)$-channel resonance/partonic
contributions. 
For instance, in this way all the data collected so far in the
charged pion electoproduction at JLAB, DESY, Cornell and CEA
can be well described~\cite{KM}.

In this work we continue our studies of the reaction $N(e,e'\pi)N'$ 
and attempt to describe the electroproduction of neutral pions above 
the resonance region. 
Recently, the $p(\gamma^*,\pi^0)p$ partial 
cross sections were measured  at JLAB
in the deeply virtual kinematics around the values of $\sqrt{s}
\simeq 2.1$~GeV  and $Q^2\simeq 2$~GeV$^2$~\cite{Collaboration:2010kna}. 
In the $\pi^0$ channel there are interesting 
features which are worth to mention. At the real photon point the $\pi^0$ 
production is probably the simplest of
all photoproduction reactions because of the limited number of allowed Reggeon
exchanges. By quantum numbers the $\gamma$-induced production of $\pi^0$ 
selects the charge conjugation $C$-parity odd configurations in the
$t$-channel.  These are the vector $\omega(782)$, $\rho(770)$ and
axial-vector $h_1(1170)$, $b_1(1235)$ Regge trajectories~\cite{Goldstein:1973xn,Yu:2011zu}.
It is also certain that  the $(\gamma,\pi^0)$ 
reaction is dominated by the $\omega(782)$-exchange.
Nevertheless, the Regge model
based on exchanges of single trajectories 
largely underestimates the $\pi^0$ electroproduction cross sections. 
The situation is very similar to the charged pion production where the Regge
model also largely underestimates the measured cross sections.
Indeed, at high values of $Q^2$ the meson-exchange contributions
are suppressed by the $\gamma \pi$-Reggeon transition form factors and
must show a rapid decrease as a function of $Q^2$.
Being dominated by the transverse $\gamma^*_{\rm T}\to\pi^0$ component,
the $\pi^0$ data, however, 
demonstrate  no pronounced $Q^2$ dependence~\cite{Collaboration:2010kna}. 
An importance of  
Regge-cut unitary corrections in this case has been demonstrated in~\cite{Laget:2010za}.
Another observable that we consider in this work is the beam single spin 
asymmetry (SSA). 
Sizeable and positive beam SSA have been found at CLAS/JLAB 
in the reaction  $p(\vec{e},e'\pi^0)p$ with the longitudinally polarized
electron beam~\cite{DeMasi:2007id}.

The outline of the manuscript is as follows. In Sec.~II  we 
briefly recall the kinematics and definitions of the cross sections in 
exclusive $\pi^{0}$ electroproduction off nucleons. 
In  Sec.~III we consider the 
contributions of vector $\omega(782)$, $\rho(770)$ and axial-vector $h_1(1170)$   and 
$b_1(1235)$ Regge trajectories. In Secs.~IV we describe the contribution of nucleon 
resonances. In Sec.~V we briefly discuss the 
$\pi^{0}$ photoproduction at forward angles. Then in Sec.~VI the model results 
are compared with the electroproduction data from JLAB. In 
Sec.~VII we calculate the cross sections in the kinematics at HERMES.   
The conclusions are summarized in Sec.~VIII.

\section{Kinematics and definitions}

At first, we recall briefly the kinematics in exclusive $\pi$ electroproduction
\begin{equation}
\label{eepireac}
e(l) + N(p) \to e'(l') + \pi(k') + N'(p'),
\end{equation}
and specify the notations and definitions of variables. 
In exclusive reaction Eq.~(\ref{eepireac}) we shall deal with an unpolarized target 
and, both unpolarized and polarized lepton beams. The differential cross 
section is given by
\begin{eqnarray}
\label{dsdte}
\frac{d\sigma}{dQ^2d\nu dt d\phi} &=&
\frac{\Phi}{2\pi} 
\left[
           \frac{d\sigma_{\rm T}}{dt}
+ \varepsilon \frac{d\sigma_{\rm L}}{dt} \right. \nonumber \\
&+& \left.  \sqrt{2\varepsilon (1+\varepsilon)}
\frac{d\sigma_{\rm LT}}{dt} \cos(\phi)  \right. \nonumber \\
&+& \left.
\varepsilon \frac{d\sigma_{\rm TT}}{dt} \cos(2\phi)
  \right. \nonumber \\
&+& \left.  h \sqrt{2\varepsilon(1-\varepsilon)}
\frac{d\sigma_{\rm LT'}}{dt} \sin(\phi)
\right],
\end{eqnarray}
where $d\sigma_{\rm T}$ is the transverse cross section, $d\sigma_{\rm L}$ is 
the longitudinal cross section, $d\sigma_{\rm TT}$ is the cross section 
originating from the interference between the transverse components of the 
virtual photon, $d\sigma_{\rm LT}$ is the cross section arising from the 
interference between the transverse and longitudinal polarizations of the 
virtual photon and $d\sigma_{\rm LT'}$ is the beam-spin polarized cross section
resulting from the interference between the transverse and longitudinal
photons and helicity $h=\pm 1$  of the incoming electron.

In the laboratory where 
the target nucleon is at rest, the $z$-axis is directed along the three 
momentum $\vec{q}=(0,0,\sqrt{\nu^2+Q^2})$ of the exchanged virtual photon 
$\gamma^*$ with $q= l-l'= (\nu,\vec{q})$, $Q^2=-q^2$, $\nu=E_e-E_e'$  and
$l(l')$ is the four momentum of incoming (deflected) electrons. In 
Eq.~(\ref{dsdte}) $\phi$ stands for the azimuthal angle between the 
electron scattering $(e,e')$ plane and $\gamma^* N \to \pi N'$ reaction 
plane. $\phi$ is zero when the pion is closest to the outgoing  
electron~\cite{Bacchetta:2004jz}. 
The differential cross section integrated over $\phi$ is denoted as
\be
\label{CSU}
\frac{d\sigma_{\rm U}}{dt}=\frac{d\sigma_{\rm T}}{dt}+\varepsilon \frac{d\sigma_{\rm L}}{dt}.
\ee

The virtual photon flux is conventionally defined as~\cite{Hand:1963bb}
\begin{equation}
\Phi = \frac{\pi}{E_e(E_e-\nu)}
\left(\frac{\alpha_e}{2\pi^2} \frac{E_e-\nu}{E_e} \frac{\mathcal K}{Q^2}
\frac{1}{1-\varepsilon}\right),
\end{equation}
with $\alpha_e \simeq 1/137$, $\mathcal K = (W^2-M^2_N)/2M_N$  and
\begin{equation}
\label{vareps}
\varepsilon = \frac{1}{ 1+2 \frac{\nu^2+Q^2}{4E_e(E_e-\nu)-Q^2}}
\end{equation}
is the ratio of longitudinal to transverse polarization of the virtual photon.

The longitudinal beam SSA in $N(\vec{e},e'\pi)N'$ scattering is defined
so that
\be
\label{BSSA}
A_{\rm LU}(\phi) \equiv
\frac{d\sigma^{\rightarrow}(\phi)-d\sigma^{\leftarrow}(\phi)}{d\sigma^{\rightarrow}(\phi)+d\sigma^{\leftarrow}(\phi)},
\ee
where $d\sigma^{\rightarrow}$ refers to positive helicity $h=+1$ of the incoming
electron. The azimuthal moment associated with the beam SSA is
given by~\cite{Bacchetta:2004jz}
\be
\label{BeamSSAmoment}
A^{\sin(\phi)}_{\rm LU} = \frac{\sqrt{2\varepsilon(1-\varepsilon)}d\sigma_{\rm
    LT'}}{d\sigma_{\rm T} + \varepsilon d\sigma_{\rm L}}.
\ee

\section{Exchange of Regge trajectories}
At first we consider the exchange of mesonic Regge trajectories.
The leading trajectories contributing to $p(e,e'\pi^{0})p$ 
are the natural parity $P=(-1)^{J}$ vector $\omega(782)$, $\rho(770)$ and the unnatural
parity $P=(-1)^{J+1}$ axial-vector $h_1(1170)$ and $b_1(1235)$ exchanges, 
with $\omega$ being the dominant trajectory. 

At high energies the differential cross section $d\sigma/dt$  in the
photoproduction $(\gamma,\pi^0)$ reaction
is characterized by pronounced dip around $-t\simeq 0.5$~GeV$^2$. 
 For instance, there is no such a dip in the $(\gamma,\pi^{\pm})$ reaction.
There exist two explanations of
this structure in $d\sigma/dt$~\cite{Collins:1980dv}. According to one
viewpoint, a coincidence of a dip with the point in $-t$ where the
$\omega$-Regge trajectory (with a slope $\alpha_{\omega}'=0.85$~GeV$^{-2}$)
\be
\label{wmegaRegge}
\alpha_{\omega}(t)\simeq 0.4+\alpha_{\omega}'t 
\ee
passes through nonsense value $-t=0.47$~GeV$^2$ demonstrates that the dips are
nonsense wrong-signature
zeros (NWSZ) of the Regge pole amplitudes.
In this case the Regge cuts are supposed to be weak and modify
the region around the dips only slightly.
The alternative viewpoint is that the Regge
trajectories are degenerated - the Regge poles are essentially featureless as
function of $-t$. Then a structure of the dip in $d\sigma/dt$ is similar to
the diffraction minima
caused by the destructive interference between the exchange of 
single Regge poles and very strong absorptive
Reggeon/Pomeron cuts in this case~\cite{Collins:1980dv}.

In electroproduction for finite $Q^2$ the two viewpoints should not necessarily
result in identical predictions~\cite{Harari:1971yt}. 
If the dip is a NWSZ its position should
remain fixed in $t$ since the $\omega$-exchange should be independent of the
external masses. On the contrary, if an absorptive cut is responsible for a
dip structure it is very likely that the dip position will move with $Q^2$ 
since the range of the interaction may well change~\cite{Harari:1971yt}. 
Interestingly, the early experiments at DESY~\cite{Berger:1977kj,Brasse:1975bg} have shown 
that the dip neither remained fixed nor moved out with $Q^2$. It simply 
disappeared as $Q^2$ was changed from its null value. This behavior has been much discussed 
in~\cite{Collins:1980dv,Nachtmann:1976be} and recently in~\cite{Laget:2010za}.

Note that, the dip problem shows up when one compares the
photoproduction data obtained at higher energies 
which indeed exhibit a pronounced dip
and then extrapolated with some
assumption to the DESY region~\cite{Berger:1977kj}. Recently, at JLAB~\cite{Dugger:2007bt} and 
ELSA~\cite{vanPee:2007tw,Bartholomy:2004uz} the $\pi^0$ 
photon-induced cross sections were measured also just above the resonance region. 
Interestingly, the dip which starts to develop only at high energies is not
there and the experimental points are only leveling off. Therefore, the dip
issue in $(\gamma^*,\pi^0)$ is rather artificial and might be related to 
the comparison of electro- and photo-cross section with different values of $W$.
 
In the following we rely on a NWSZ scenario. In this case the  
Regge propagator of the $J=1$ mesons reads
\ba
\label{ReggeProp}
D_{\mathcal{R}}^{\mu\nu} &=&
\left(-g^{\mu\nu}+\frac{k^{\mu}k^{\nu}}{m_{\mathcal{R}}^2} \right) 
 \left[\frac{1 - e^{-i\pi\alpha_{\mathcal{R}}(t)}}{2}\right]
\nonumber \\
&\times&
\left(- \alpha'_{\mathcal{R}} \right) \Gamma[1-\alpha_{\mathcal{R}}(t)]
e^{\ln(\alpha'_{\mathcal{R}}s)(\alpha_{\mathcal{R}}(t)-1)} ,
\ea
where $t=k^2$ and $k=k'-q=p-p'$. 
In (\ref{ReggeProp}) the $\Gamma$-function contains the pole propagator 
$\sim 1/\sin(\pi\alpha_{\mathcal{R}}(t))$ but no zeroes and the amplitude 
zeroes only occur through the factor 
$1 - e^{-i\pi\alpha_{\mathcal{R}}(t)}$. When using Eq.~(\ref{ReggeProp}) 
the region around the dip will remain strongly underestimated.
If one assumes that the cross
section at the dip is dominated by unnatural parity exchanges then the
observed dip must be filled up by the axial-vector $b_1(1235)$ and/or
$h_1(1170)$ Regge trajectories. However, this
assumption is in contradiction to the experimental data. The energy dependence
of the dip is approximately that of the region outside the dip.

A possible way to fill the dip region is to allow
absorption cuts in the model (see Ref.~\cite{Sibirtsev:2009kw} for the recent calculations and
references therein). A somewhat different prescription has been used in~\cite{Guidal:1997hy}.
Nevertheless, as we shall see in the following in the model without strong cuts  there is 
an alternative mechanism which allows to fill the region around the dip. Furthermore, we will
show that this mechanism is at the origin of the large transverse cross section observed in
the $\pi^0$ electroproduction.

\begin{widetext}
\subsection{Vector $\omega(782)$ and $\rho(770)$ exchange currents}
In the following we give the expressions for the currents (amplitudes) only. 
Then the L/T separated virtual-photon nucleon 
cross sections are calculated using the expressions given in the Appendix~A of Ref.~\cite{KM}.

The currents $J_{V}^{\mu}$ describing the exchange of 
$V=\rho$- and $\omega$-Regge trajectories in the reaction $p(\gamma^*,\pi^0)p$ are
given by

\begin{eqnarray}
\label{rho}
-iJ_{V}^{\mu}(\gamma^*p\to\pi^0p)
= - i
\
G_{V\gamma\pi}
\, G_{V NN}
F_{V\gamma\pi}(Q^2)
{\varepsilon^{\mu\nu\alpha\beta} q_{\nu}
 k_{\alpha}}  
\bar{u}_{s'}(p')
\left[  (1 + \kappa_{V}) \gamma_{\beta}
- \frac{\kappa_{V}}{2M_p} (p+p')_{\beta} \right] u_s(p) \nonumber \\
\times \left[\frac{1 - e^{-i\pi\alpha_{V}(t)}}{2}\right]
\left(- \alpha'_{V} \right) \Gamma[1-\alpha_{V}(t)]
e^{\ln(\alpha'_{V}s)(\alpha_{V}(t)-1)},
\end{eqnarray}
\end{widetext}
In the $V=\rho(770)$ case the $G_{\rho NN}=3.4$ and $\kappa_{\rho}=6.1$ are the vector and 
anomalous tensor coupling constants, respectively. The $\rho$-trajectory 
adopted here reads 
\be
\label{rhoRegge}
\alpha_{\rho}(t) = 0.53+\alpha_{\rho}'t
\ee
with a slope $\alpha_{\rho}'=0.85$~GeV$^{-2}$.  
For the transition form factor $F_{\rho\gamma\pi}(Q^2)$ we use
$F_{\rho\gamma\pi}(Q^2)=(1+Q^2/\Lambda^2_{\rho\gamma\pi})^{-1}$ with 
$\Lambda_{\rho\gamma\pi}=m_{\omega(782)}$.
All these parameters and Eq.~(\ref{rho}) (up to the isospin factor) are  the
same as have been used in the charged pion electroproduction of
Ref.~\cite{KM}. 

In the natural parity sector 
the dominant contribution comes from the exchange
of the $\omega(782)$-Regge trajectory given by Eq.~(\ref{wmegaRegge}). 
From the fit to high energy photoproduction data we obtain $G_{\omega
  NN}=18\pm 1$.
We also neglect the anomalous
tensor component $\kappa_{\omega}\simeq 0$~\cite{Guidal:1997hy}.
 For the transition form factor $F_{\omega\gamma\pi}(Q^2)$ we use
a monopole form factors
$F_{\omega\gamma\pi}(Q^2)=(1+Q^2/\Lambda^2_{\omega\gamma\pi})^{-1}$ with the
cut-off $\Lambda_{\omega\gamma\pi}$ being a fit parameter.

The ${V\gamma\pi}$ coupling constants $G_{V\gamma\pi}$ can be deduced
from the radiative $\gamma\pi$ decay widths of $V$:
\be
\label{Vgammapi}
\Gamma_{V \to \gamma \pi^{0}} = \frac{\alpha_e}{24}
\frac{G_{V \gamma\pi}^2}{m_{V}^3} 
\left({m_{V}^2} - {m_{\pi}^2}\right)^3.
\ee
The measured widths are~\cite{PDG}
$
\Gamma_{\rho^{0}\to \gamma\pi^{0}}=(89 \pm 11)\mbox{~keV}$ and 
$\Gamma_{\omega\to \gamma\pi^{0}}=(764 \pm 51)\mbox{~keV}$
where the central values correspond to $G_{\rho\gamma\pi} = 0.84$~GeV$^{-1}$,
$G_{\omega\gamma\pi} = 2.4$~GeV$^{-1}$.

\begin{widetext}
\subsection{Axial-vector $b_1(1235)$ and $h_1(1170)$ exchange currents}
We further take into account 
the exchanges of lowest lying $C$-parity odd $J^{PC}=1^{+-}$ 
axial-vector $b_1(1235)$ and $h_1(1170)$ mesons with
$I^G=1^+$ and $I^G=0^-$, respectively. The results reported
in~\cite{DeMasi:2007id} suggest an important role of axial-vector mesons in
the description of polarization observables.
Together with $h_1(1380)$ state
$(I^G=0^-)$ they belong to the nonet  of $C=-1$ axial-vector mesons. Assuming
the ideal mixing pattern between the singlet and $I=0$ of the octet the $h_1(1380)$ meson
decouples from the $\pi^0\gamma$ channel. The existing empirical information~\cite{PDG} does
not allow to fix the coupling constant of $h_1(1170)$ to
$\pi^0\gamma$ channel. From mere SU(3) arguments the width of $h_1(1170)$ into
the $\pi^0\gamma$ decay channel should be about an oder of magnitude larger than the decay
width of  $b_1(1235) \to \pi^0\gamma$.  However, the strength of their interaction with nucleons can be
estimated only indirectly. For instance, in the analysis of charged pion production we
did not find any strong evidence for the contribution of $b_1(1235)$ trajectory~\cite{KM}.
 Since, the Lorentz structure of the
interaction vertices are essentially the same, both trajectories can be combined in a single
current describing the exchange of axial-vector mesons.

The hadronic current $-iJ_{B}^{\mu}$ describing the exchange of $B=b_1(1235)$
and $h_1(1170)$ Regge trajectories takes the form~\cite{KM}
\begin{eqnarray}
\label{Jaxial}
-iJ_{B}^{\mu}(\gamma^*p\to\pi^0p)
&=& 
G_B
\Big[  k^{\mu} q^{\nu} -(qk)g^{\mu\nu}\Big] (p+p')_{\nu} 
\bar{u}_{s'}(p') \gamma_{5} u_s(p) \nonumber \\
& \times & \left[\frac{1 - e^{-i\pi\alpha_{B}(t)}}{2}\right]
\left(- \alpha'_{B} \right) \Gamma[1-\alpha_{B}(t)]
e^{\ln(\alpha'_{B}s)(\alpha_{B}(t)-1)}.
\end{eqnarray}
The effective coupling constant $G_B\sim $~GeV$^{-2}$ absorbs the couplings in the
$B\gamma\pi^0$ and $BNN$ vertices. 
The $\pi$ and $b_1(1235)$ Regge trajectories 
 are nearly degenerate. 
Therefore, we assume $\alpha_{h_1}(t)\simeq \alpha_{b_1}(t)=\alpha_{\pi}(t)$
where $\alpha_{\pi}(t)=0.74 (t-m_{\pi}^2)$ is the $\pi$-trajectory.
In the $B\gamma^*\pi$ vertex we use the monopole form factor 
$F_{B\gamma\pi}(Q^2) = (1+Q^2/\Lambda_{B\gamma\pi}^2)^{-1}$ with
$\Lambda_{B\gamma\pi}=m_{V}$ where $m_{V}$ is, {\it e.g.},  an average mass of
vector $\omega(\rho)$ mesons. Note that,
Eq.~(\ref{Jaxial}) does not interfere with the amplitude describing the
exchange of vector mesons, see Eq.~(\ref{rho}). That is, in the interference
term the trace over the spinor indices involve the $\gamma_5$ with odd number of $\gamma$ matrices.

\section{Effect of nucleon resonances}
So far we have considered the $t$-channel meson-exchange contributions. In this section we
estimate the residual effect from the propagation of the nucleon and its
excitations in the the $s$- and $u$-channels. 
We start from the nucleon Born terms in the $s$-
and $u$-channels. The contribution (the sum) of the corresponding $s$- and
$u$-channel  diagrams in the reaction $p(\gamma^*,\pi^0)p$ 
takes the form~\cite{Kaskulov:2008ej}
\ba
\label{PipolePl}
-iJ^{\mu}_s-iJ^{\mu}_u &=&  g_{\pi NN} \,
\bar{u}_{s'}(p')
\left[
\mathcal{F}_{s}(Q^2,s,t)
\gamma_5\frac{(p+q)_{\sigma}\gamma^{\sigma}\gamma^{\mu} + M_p\gamma^{\mu}}
{s-M_p^2+i0^+}  \right. \nonumber \\ 
&+& \mathcal{F}_{u}(Q^2,u,t)
\frac{ \gamma^{\mu}(p'-q)_{\sigma}\gamma^{\sigma} + M_{p}\gamma^{\mu}
}{u-M_p^2+i0^+}\gamma_5 
- \left.
\gamma_5 [\mathcal{F}_{s}(Q^2,s,t)-\mathcal{F}_{u}(Q^2,u,t)] \frac{(k-k')^{\mu}}{Q^2}
\right] u_s(p), 
\ea
where $\mathcal{F}_{s(u)}(Q^2,s(u),t)$ stands for the proton $s(u)$-channel
transition form factors. In Eq.~(\ref{PipolePl}) $g_{\pi NN}=13.4$
is the pseudoscalar $\pi N$ coupling constant.
In the sum of two Born amplitude the
current conservation condition, $q_{\mu}J^{\mu}_{s+u} =0$, is satisfied in the
presence of different form factors, $\mathcal{F}_{s}$ and
$\mathcal{F}_{u}$, which in general can depend on values of
$(Q^2,s(u),t)$. Eq.~(\ref{PipolePl}) has been obtained using the
requirement that the modified electromagnetic vertex functions  entering the 
amplitude obey the same Ward-Takahashi identities as the bare 
ones, see~\cite{Kaskulov:2008ej} and references therein.

The nucleon resonances are taken into account using the Bloom-Gilman duality
connection~\cite{Bloom:1971ye,Bloom:1970xb,Elitzur:1971tg} between the 
resonance form factors and deep inelastic structure functions.
This allows to absorb the contribution of an infinite sum of resonances into the
generalized form factors $\mathcal{F}_{s(u)}(Q^2,s(u),t)$. 
The details concerning the dual connection used here are
described in Ref.~\cite{KM}. Here we only give an explicite expressions for
the $s$- and $u$-channel form factors. They are given by
\be
\label{F1_s_res}
{F}_{s}(Q^2,s) = 
\frac{\ds
s \ln\left[\frac{ \xi Q^2}{M_p^2}+1\right] 
\frac{ (2\xi Q^2+s)}{(\xi Q^2)^2} 
- \frac{s (\xi Q^2+s) }{ \xi Q^2 \,(\xi Q^2+M_p^2)} 
+ \ln\left[
  \frac{s-M_p^2}{M_p^2}\right] -i\pi}
{\ds \left(\frac{\xi Q^2}{s}+1\right)^2 \left( \frac{
    s^2+2s M_p^2}{ 2 M_p^4} + \ln\left[\frac{ s-M_p^2}{ M_p^2}\right]-i\pi \right)},
\ee
\be
\label{F1_u_res}
{F}_{u}(Q^2,u) = 
\frac{\ds
u \ln\left[\frac{ \xi Q^2}{M_p^2}+1\right] 
\frac{ (2\xi Q^2+u)}{(\xi Q^2)^2} 
- \frac{u (\xi Q^2+u) }{ \xi Q^2 \, (\xi Q^2+M_p^2)} 
+ \ln\left[
  \frac{M_p^2-u}{M_p^2}\right]}
{\ds \left(\frac{\xi Q^2}{u}+1\right)^2 \left( \frac{
    u^2+2u M_p^2}{ 2 M_p^4} + \ln\left[\frac{M_p^2-u}{ M_p^2}\right]\right)},
\ee
\end{widetext}
with the cut off $\xi=0.4$~\cite{KM}.
Note that this parameter is fixed in the charged pion
electroproduction and we do not refit it in the neutral pion channel.
Furthermore, for generalized form factors in the $s$- and $u$-channels 
we take the same phase as determined in the charged pion production,
that is
\be
\label{F1generic}
\mathcal{F}_{s(u)}(Q^2,s(u),t) = {F}_{s(u)}(Q^2,s(u)) (t-m_{\pi}^2) 
 \mathcal{R}(\alpha(t)).
\ee
where ${F}_{s(u)}$ are given by Eqs.~(\ref{F1_s_res})
and~(\ref{F1_u_res}) and the Regge factors $\mathcal{R}_{s(u)}$ are given by 
\ba
\mathcal{R}_{s}(\alpha(t)) = {e^{-i\pi\alpha(t)}}   
\left(- \alpha' \right) \Gamma[-\alpha(t)]
e^{\alpha(t) \ln(\alpha's)}, \nonumber \\ 
\mathcal{R}_{u}(\alpha(t)) =  
\left(- \alpha' \right) \Gamma[-\alpha(t)]
e^{\alpha(t) \ln(\alpha's)}, \nonumber
\ea
where $\alpha(t)=\alpha'(t-m_{\pi}^2)$, $\alpha'=\alpha_{\pi}'/(1+2.4
Q^2/W^2)$ and $\alpha_{\pi}'=0.74$ stands for the slope of the $\pi$-Regge trajectory.

\begin{figure*}[t]
\begin{center}
\includegraphics[clip=true,width=2\columnwidth,angle=0.]
{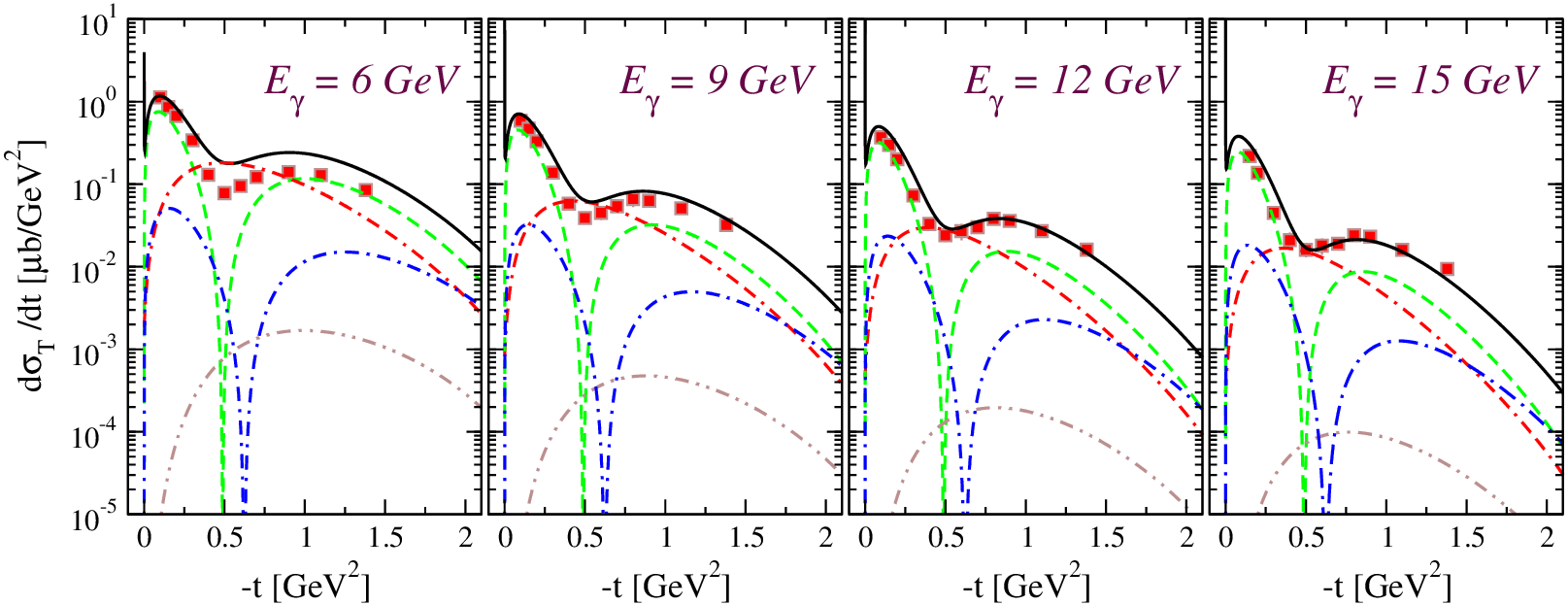}
\caption{\label{Pi0Photo} 
(Color online) Differential cross section $d\sigma_{\rm
  T}/dt$ in the reaction $p(\gamma,\pi^0)p$ at high energies. Different plots correspond to
different values of photon
energy
$E_{\gamma}=6$, 9, 12 and 15~GeV in the laboratory.  The experimental data are
from~\cite{Anderson:1971xh}.
The solid curves describe
the model results. The dashed and dash-dotted curves describe the exchange of
$\omega(782)$ and $\rho(770)$ Regge trajectories, respectively. The
dash-dash-dotted curves that fill the region around the dip describe the $s$-
and $u$- channel resonance contributions. The contribution of the axial-vector mesons
$b_1(1230)$ and $h_1(1170)$ is shown by the dot-dot-dashed curves.
The strong rise of the cross section at
extreme forward angles is due to the Primakoff effect.
\vspace{-0.5cm}
}
\end{center}
\end{figure*}

\section{Photoproduction}
In this section we consider briefly the reaction
$p(\gamma,\pi^0)p$ at the real photon point. 
In Fig.~\ref{Pi0Photo} we compare the model results with photoproduction 
data from~\cite{Anderson:1971xh} for values of $E_{\gamma}=$ 6, 9, 12 and 15~GeV in the
laboratory. 
The dashed and dash-dotted curves in Fig.~\ref{Pi0Photo} describe the contributions 
of the $\omega$- and $\rho$-Regge trajectories, respectively. 
Here the value of $G_{\omega NN}=18$ has been used.  
Within the Regge phenomenology  with NWSZ the contributions from
$\omega(782)$ and $\rho(770)$ trajectories disappear around the value of $-t\simeq
0.5$~GeV$^2$. Indeed, using the linear Regge trajectories $\alpha_{\omega}(t)$
and $\alpha_{\rho}(t)$ adopted here, see Eqs.~(\ref{wmegaRegge}) and~(\ref{rhoRegge}), the
zeroes occur at $-t_{\omega}=0.47$~GeV$^2$ and $-t_{\rho}=0.62$~GeV$^2$, correspondingly.
Therefore, the observed dip near $-t\simeq 0.5$~GeV$^2$ can be readily
explained. However, the region around the dip remains strongly underestimated in this case.

We further add to the Regge amplitudes the amplitude resulting from the $s$- and
$u$-channel contributions, see Eq.~(\ref{PipolePl}). 
Individually the $s(u)$-channel terms are large like in the $\pi^{\pm}$
production at forward angles, see Ref.~\cite{KM}. 
However, in the neutral pion channel the sum of $s$- and
$u$-channel pole terms partially cancel. For instance, at forward angles the
sum of pole terms practically cancel and the dash-dash-dotted 
curves in Fig.~\ref{Pi0Photo} result from the partial cancellation of 
two contributions. In the following we rely on this scenario where the 
$s(u)$-channel pole terms explain the region around 
the dip. This is in line  with the results of~\cite{LR} where the nucleon resonance 
background has been considered as the main dip filling mechanism.

The solid curves in Fig.~\ref{Pi0Photo} describe the sum of all the production 
amplitudes. The very strong rise of the cross section at extreme forward
directions is due to the Primakoff effect~\cite{Pirmakoff:1951pj}. The
Primakoff amplitude is given in Ref.~\cite{Kaskulov:2011ab} and we do not 
repeat these formulae here. Note that, the dominance of Regge contributions and
the cancellation of $s(u)$-channel pole terms at extreme forward angles
simplifies the analysis of Primakoff $\pi^0$ data off nuclei~\cite{Kaskulov:2011ab}.

\begin{figure}[b]
\begin{center}
\includegraphics[clip=true,width=1\columnwidth,angle=0.]
{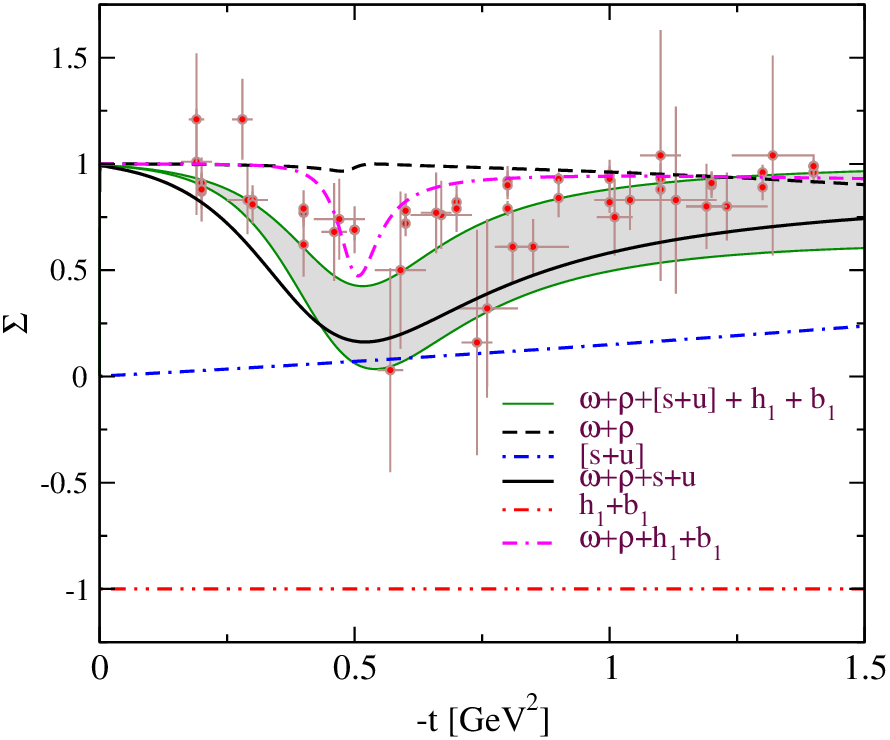}
\caption{\label{Pi0photoAsymmetry} 
(Color online) Polarized photon asymmetry $\Sigma$ in the reaction $p(\gamma,\pi^0)p$. The
compilations of
experimental data with $E_{\gamma}=2.5\div 10$~GeV 
are from~\cite{Anderson:1971xh,Bellenger:1969aa}. The dashed curve curve
describe the asymmetry generated by the $\omega$- and $\rho$-Regge
trajectories. The addition of $s(u)$-channel pole terms (dash-dotted curve) 
results in $\Sigma$ shown by solid curve. The dash-dotted curve describe the
combined effect of vector
$\omega(\rho)$ and axial-vector $h_1(b_1)$ exchanges (dot-dot-dashed line). The shaded region
describe the model results with couplings of the axial-vector mesons as described
in the text.
\vspace{-0.3cm}
}
\end{center}
\end{figure}

In Fig.~\ref{Pi0photoAsymmetry} we show our results for the polarized photon 
asymmetry $\Sigma$ defined as
\begin{equation}
\Sigma = \frac{d\sigma_{\perp} - d\sigma_{\parallel}}
{d\sigma_{\perp} + d\sigma_{\parallel}}.
\end{equation}
Here $d\sigma_{\perp}$ and $d\sigma_{\parallel}$ are the differential cross
sections where the incoming photons are polarized in the direction perpendicular
and parallel to the $\gamma\pi^0$ reaction plane, respectively. With our 
definition of partial L/T cross sections  the asymmetry is 
$\Sigma = - \frac{d\sigma_{\rm TT}}{d\sigma_{\rm T}}$. This observable is
known to be sensitive to the reaction mechanism around the dip. In
Fig.~\ref{Pi0photoAsymmetry} we also plot the compilation of high 
energy data~\cite{Anderson:1971xh,Bellenger:1969aa} in the range 
$E_{\gamma}=2.5\div 10$~GeV in the laboratory. 
In a model with $\omega$- and $\rho$-Regge trajectories only (dashed
curve) this observable is essentially unity. Because of the weak energy 
dependence of the model results the calculations are performed for 
the average value of $E_{\gamma}=4$~GeV. The asymmetry generated by the $s$- and
$u$-channel contributions only is shown by the dash-dotted line.
The solid curve is a combined effect of the $\omega(\rho)$-Regge and the 
$s(u)$-channel components of the cross section. As one can see, the addition of the $s$- and
$u$-channel pole terms results in an asymmetry around the dip which is qualitatively in
agreement with data. 

In the model without $s(u)$-channel effects the polarized photon 
asymmetry $\Sigma$ could be generated by the axial-vector mesons. 
In the model with axial-vector mesons only the asymmetry 
is $\Sigma=-1$ (dot-dot-dashed line in Fig.~\ref{Pi0photoAsymmetry}). 
The dash-dash-dotted curve in Fig.~\ref{Pi0photoAsymmetry} 
describe the Regge model with the vector and axial-vector exchanges. 
In these calculations we used the values of $G_B=\pm 11$~GeV$^{-2}$.
Note that, with proper $BNN$ interactions constrained by 
$C$-parity~\cite{KM} the $C$-parity odd axial-vector and vector exchanges do not interfere
with each other. Therefore, this result does not depend on the sign of the
$G_B$ coupling in Eq.~(\ref{Jaxial}). 

The shaded band in Fig.~\ref{Pi0photoAsymmetry} describes the model results 
with vector, axial-vector ($G_B=\pm 11$~GeV$^{-2}$) and $s(u)$-channel
contributions. The axial-vector currents have a potential to improve our description of
photon asymmetry data by their interference with
$s(u)$-channel contributions. 
As one can see, however, in Fig.~\ref{Pi0photoAsymmetry}, the contribution 
of axial vector-mesons (dot-dot-dashed curves) is marginally small and can be readily neglected 
in the rest of observables studied here.

\section{Electroproduction: JLAB data}
\begin{figure}[t]
\begin{center}
\includegraphics[clip=true,width=1\columnwidth,angle=0.]
{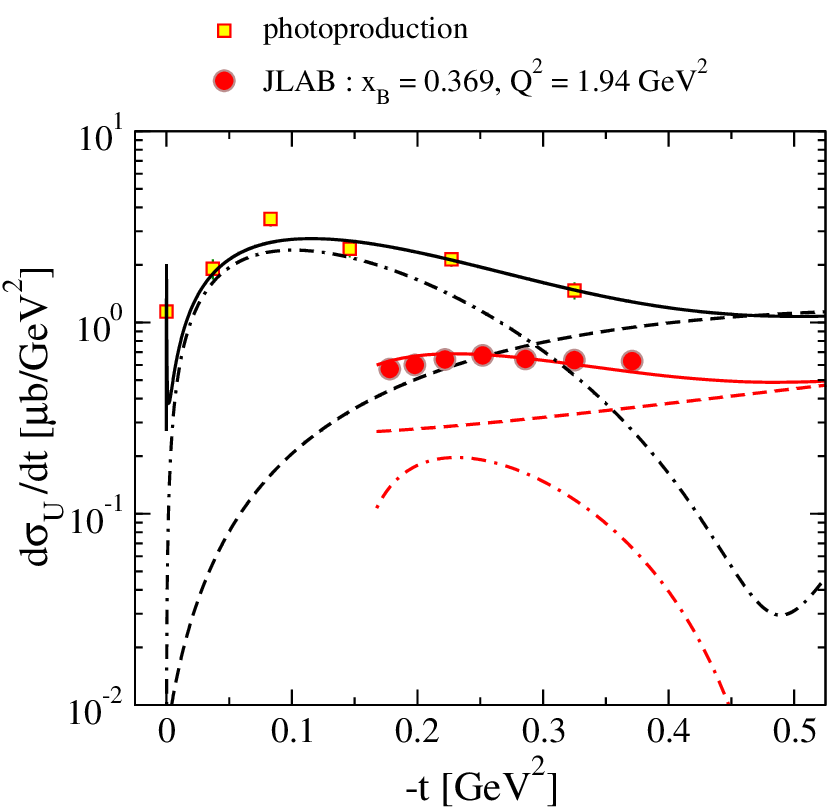}
\caption{\label{Pi0PhotoElectro} 
(Color online) Comparison of the differential cross sections $d\sigma_{\rm T}/dt$ and
$d\sigma_{\rm U}/dt=d\sigma_{\rm T}/dt + \varepsilon d\sigma_{\rm L}/dt$ in $\pi^0$
photoproduction and electroproduction off protons, respectively. The upper
 solid, dashed
and dash-dotted curves (black) belong to the photoproduction reaction. The model
results are described by the solid curves and include the Regge-exchange
(dash-dotted) and
resonance (dashed curves) contributions. The photoproduction (filled squares) and
electroproduction (filled circles) data are from~\cite{Braunschweig:1970dp} and 
\cite{Collaboration:2010kna}, respectively.}
\vspace{-0.3cm}
\end{center}
\end{figure}

In Fig.~\ref{Pi0PhotoElectro} 
we compare our results in the reactions 
with real $p(\gamma,\pi^0)p$ and virtual
$p(\gamma^{*},\pi^0)p$ photons. 
The forward photo-data (filled squares) with $E_{\gamma}=3$~GeV are from~\cite{Braunschweig:1970dp}. 
The electroproduction
data~\cite{Collaboration:2010kna} (filled circles) contain both the transverse
and longitudinal components, that is
$d\sigma_{\rm U}/dt=d\sigma_{\rm T}+\varepsilon d\sigma_{L}/dt$ and correspond to
values of $Q^2=1.94$~GeV$^2$, $x_{\rm B}=0.37$ and $E_e=5.75$~GeV.
Again in photoproduction at forward angles the Regge-exchange contributions (upper
dash-dotted curve) dominate the
cross section.  The Primakoff effect is relevant at extreme forward angles 
and is anyway cut by $-t_{min}$ in electroproduction.
The resonances (upper dashed curve) take over in the region of the NWSZ dip.
However, in the electroproduction of $\pi^0$ the situation is 
different - at least in the context of the present model.
Neglect the contribution of axial-vector mesons. 
Then the only model parameter to be fitted to the
electroproduction data
is the cut off in the form factor describing the $\gamma^*\omega\pi^0$
vertex. The calculation presented in  Fig.~\ref{Pi0PhotoElectro} correspond 
to $\Lambda_{\gamma\omega\pi^0}=1.2$~GeV.
In electroproduction with virtual photons the
meson-exchange currents (dash-dotted curve) get largely reduced as compared to the
resonance contributions (dashed curve). Because of the weaker $Q^2$
dependence, the latter dominate in  electroproduction
and  by the interference with Regge exchanges explain the cross section
$d\sigma_{\rm U}/dt$ observed in the experiment. A model with only the
Regge-exchange amplitudes underestimates the cross section by about one order
of magnitude.

\begin{figure*}
\begin{center}
\includegraphics[clip=true,width=1.6\columnwidth,angle=0.]
{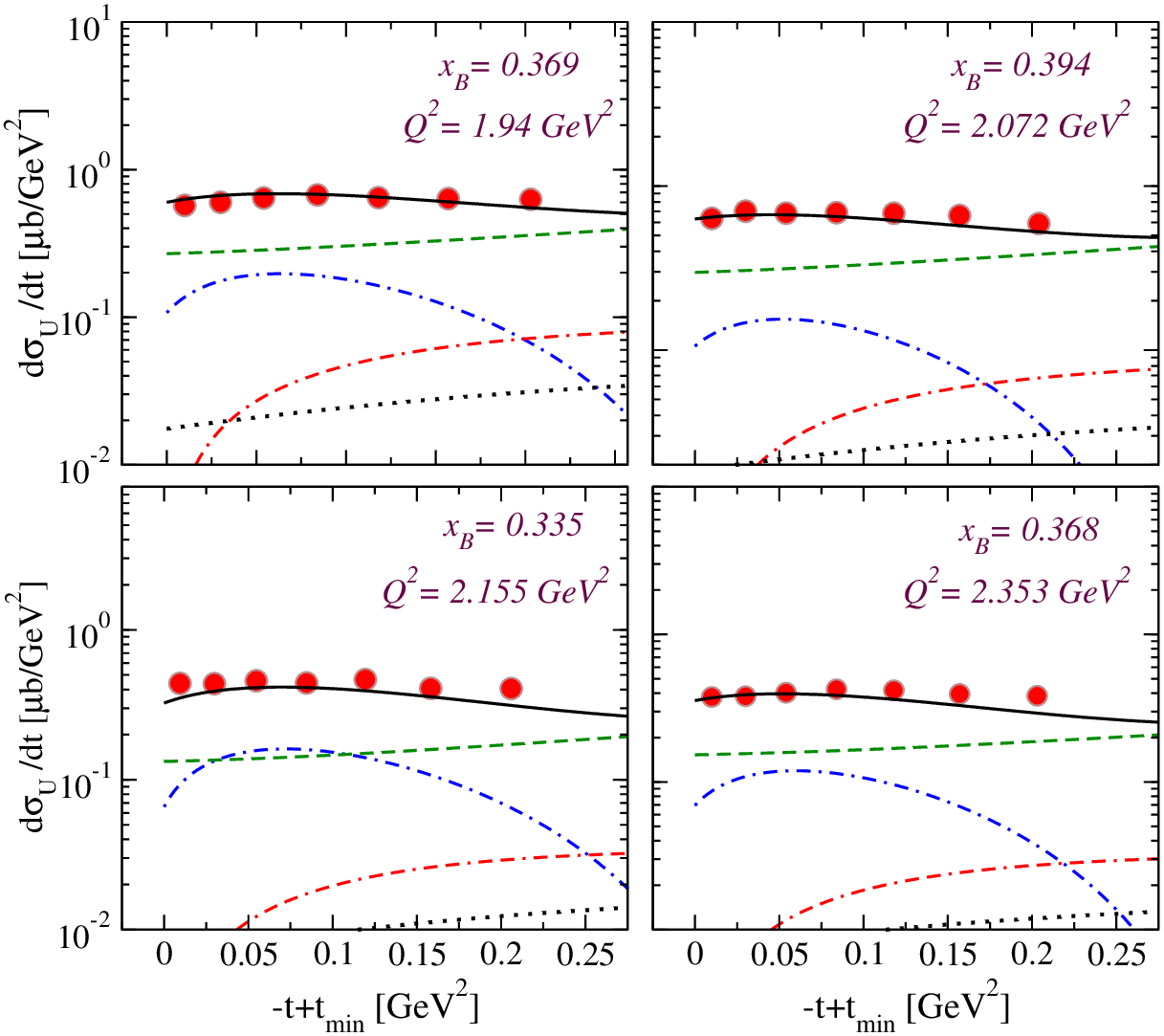}
\caption{\label{Pi0JLABkin} 
(Color online) $-t+t_{min}$ dependence of 
$d\sigma_{\rm U}/dt=d\sigma_{\rm T} + \varepsilon d\sigma_{\rm L}/dt$ differential cross section 
in exclusive reaction $p(\gamma^*,\pi^0)p$. 
The experimental data are from~\cite{Collaboration:2010kna}
The numbers displayed in the plots are the average values of $(Q^2,x_B)$
for each bin. 
The solid curves are the model results and include the Regge and nucleon
resonance contributions. The dash-dotted curves describe the
Regge exchange contributions and the dashed curves describe the effect of nucleon
resonances. The dotted curves are the contribution of the nucleon pole terms.
The dash-dash-dotted curves describe the contribution of the longitudinal
component $\varepsilon d\sigma_{\rm L}/dt$ to the total unseparated cross
section $d\sigma_{\rm U}/dt$. 
\vspace{-0.5cm}
}
\end{center}
\end{figure*}

\begin{figure*}
\begin{center}
\includegraphics[clip=true,width=1.6\columnwidth,angle=0.]  
{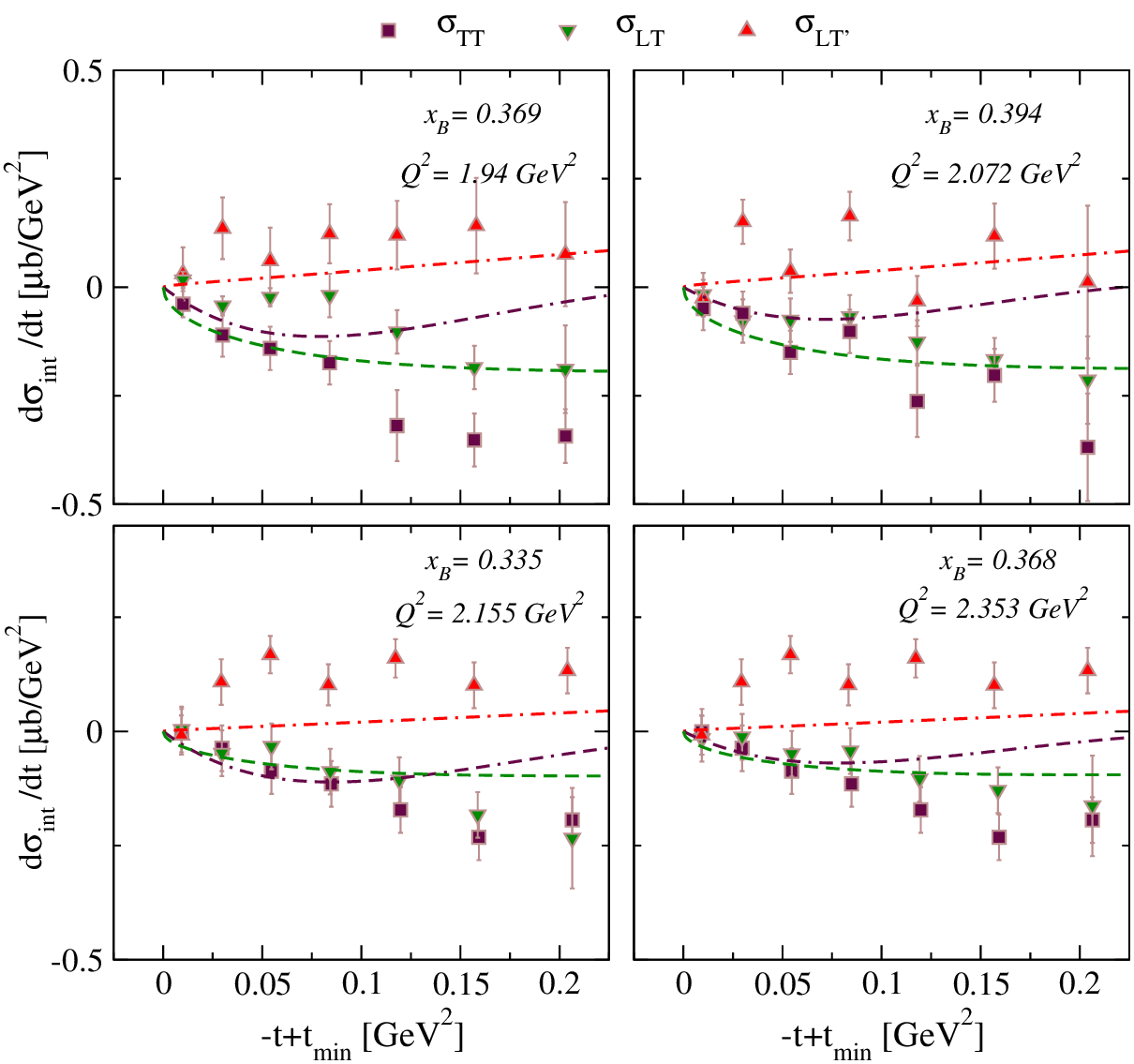}
\caption{\label{Pi0JLABkinInt} 
(Color online) $-t+t_{min}$ dependence of the interference 
$d\sigma_{\rm LT}/dt$, $d\sigma_{\rm TT}/dt$ and $d\sigma_{\rm LT'}/dt$ differential cross sections 
in exclusive reaction $p(\gamma^*,\pi^0)p$. The experimental data are 
from\cite{Collaboration:2010kna}. The dashed, dash-dash-dotted and dash-dotted curves
are the model results for  $d\sigma_{\rm LT}/dt$, $d\sigma_{\rm TT}/dt$ and
$d\sigma_{\rm LT'}/dt$, respectively.
\vspace{-0.7cm}
}
\end{center}
\end{figure*}

In Fig.~\ref{Pi0JLABkin} we show the $Q^2$ dependence of $d\sigma_{\rm U}/dt$
as a function of $-t+t_{min}$ 
for four values of $Q^2= 1.94$~GeV$^2$, 2.07~GeV$^2$, 2.16~GeV$^2$ and
$2.35$~GeV$^2$ and $x_{\rm B}=0.37$, 0.39, 0.34 and 0.37,
respectively. The experimental data are
from~\cite{Collaboration:2010kna}. The solid curves describe the model results
and include the $\omega(\rho)$-Regge and $s(u)$-channel resonance
contributions. The dash-dotted
curves which decrease rapidly as a function of $-t$ 
correspond to the Regge-exchange contributions only. On the contrary, the
resonance contributions (dashed curves) are practically flat and show no
pronounced $Q^2$ behavior. 
This behavior can be traced back to the Bloom-Gilman duality connection which
demands the hardening of the higher lying resonance transition form factors as a
function of $Q^2$~\cite{Elitzur:1971tg}.
Qualitatively it is also similar to the behavior of
charged pion electroproduction cross section~\cite{KM}. 
It is interesting to see the
contribution of resonances as compared to the contribution of the nucleon pole
terms. Replacing the resonance form factors in Eqs.~(\ref{F1_s_res})
and~(\ref{F1_u_res}) by the nucleon form
factors $F_{s(u)}\to F_{1}^p$ one gets the contribution of $s$- and $u$-channel nucleon Born
terms (dotted curves). The parameterization of $F_{1}^p$ used here is 
given in~\cite{KM} and follows the results
of~\cite{Kaskulov:2003wh}.  The
difference between the dashed and dotted curves is the effect of nucleon
resonances. As one can see, the excitation of resonances largely 
dominates the electroproduction cross section.

In Fig.~\ref{Pi0JLABkin}
the dash-dash-dotted curves describe the contribution of the longitudinal 
cross section $\varepsilon d\sigma_{L}/dt$  to the total unseparated 
cross section $d\sigma_{\rm U}/dt$. The impact of
$d\sigma_{\rm L}/dt$ is marginal and the observed cross section is totally
transverse. Furthermore, we find that the longitudinal cross section
$d\sigma_{\rm L}/dt$ decreases as a function
of $Q^2$. This is in agreement with the two-component model of Ref.~\cite{Kaskulov:2008xc}.
However, this is at variance with the present expectations based on pQCD
models which rely on the dominance of the longitudinal cross section $d\sigma_{\rm L}/dt$.
To check this result against the data one needs the 
experimental Rosenbluth  separation of $d\sigma_{\rm U}/dt$ into 
its $d\sigma_{\rm T}/dt$  and $d\sigma_{\rm  L}/dt$ components. 

In Fig.~\ref{Pi0JLABkinInt} we compare the model results with the interference cross section
measured at JLAB~\cite{Collaboration:2010kna}. Note that,
in~\cite{Collaboration:2010kna} the $\pi^0$ data have been presented using different
conventions, see Eq.(\ref{dsdte}), for the partial cross sections. 
The interference term ${d\sigma_{\rm TT}}/{dt}$ 
and ${d\sigma_{\rm U}}/{dt}$ are the same in both conventions.
$d\sigma_{\rm LT}/{dt}$ and $d\sigma_{\rm
    LT'}/{dt}$  in Fig.~\ref{Pi0JLABkinInt} are plotted according to~\cite{Collaboration:2010kna}. 
Our calculations are in qualitative
agreement with data and at forward angles result in positive values of $d\sigma_{\rm LT'}/dt$ and
negative values of $d\sigma_{\rm TT}/dt$ and $d\sigma_{\rm LT}/dt$
interference cross section. For instance, the model of
Ref.~\cite{Laget:2010za} agrees with the sign of $d\sigma_{\rm LT'}/dt$ and
$d\sigma_{\rm TT}/dt$ cross sections and predicts the positive values of $d\sigma_{\rm
  LT}/dt$. Interestingly, the preliminary data from Hall C at JLAB~\cite{Kubarovsky:2008rs}
show also the positive $d\sigma_{\rm LT}/dt$ at
least at higher values of $-t$. Therefore, both the experimental and theoretical
results concerning the sign of $d\sigma_{\rm LT}/dt$ are still not settled.

\begin{figure*}
\begin{center}
\includegraphics[clip=true,width=1.9\columnwidth,angle=0.]
{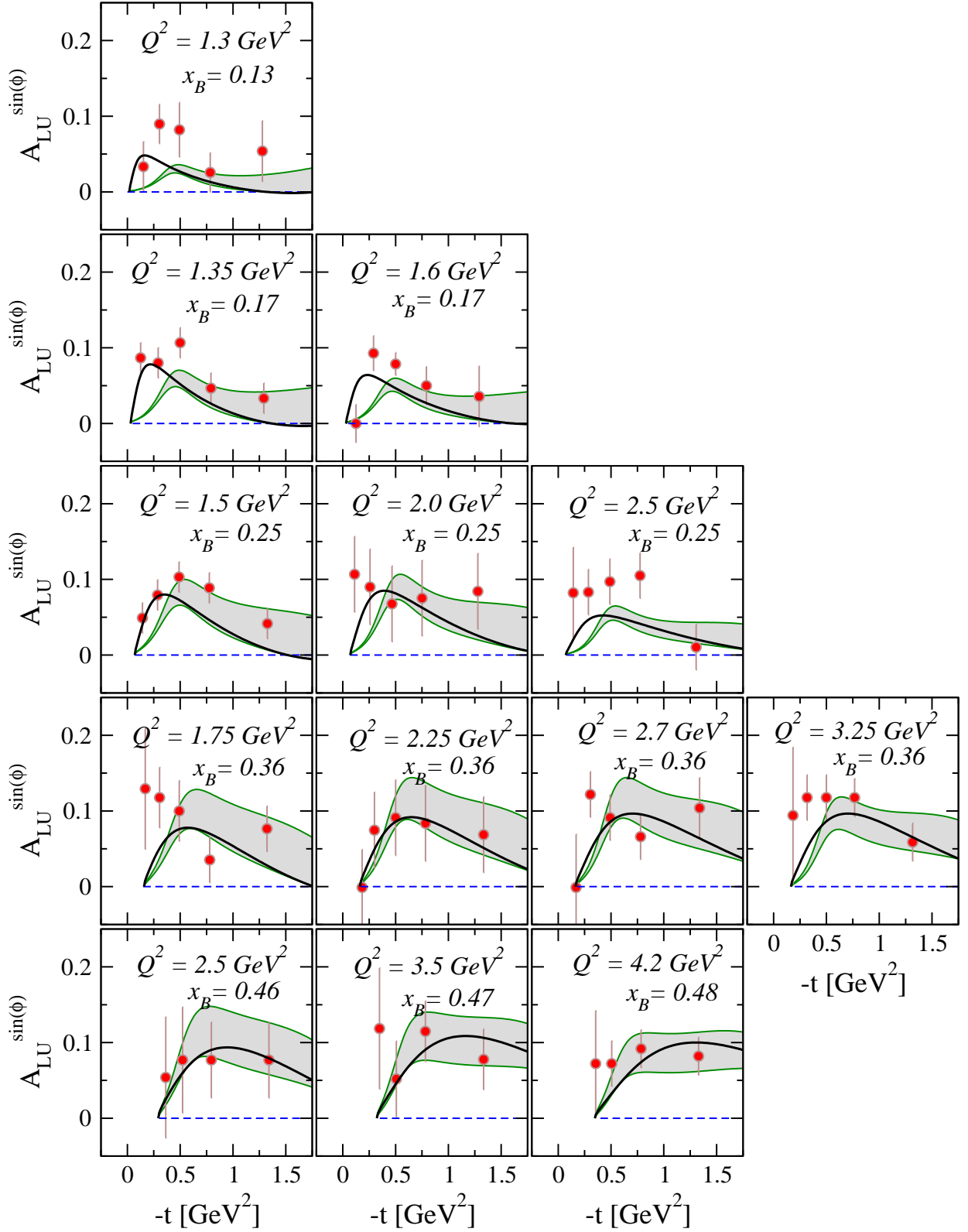}
\caption{\label{BeamSpinAsymmPi0} 
(Color online) The beam spin azimuthal moment
$A^{\sin(\phi)}_{\rm LU}$ in exclusive
reaction $p(\gamma^*,\pi^0)p$ as a function of $-t$. The experimental data are
from~\cite{DeMasi:2007id}.  The dashed curves describe 
the results (the asymmetry is zero) based on exchange of
$\omega,\rho,h_1$ and $b_1$ Regge trajectories 
(without the resonance contributions). The solid curves are the model results which account for the
effect of nucleon resonances only. The shaded bands describe the calculations
with Regge and resonances contributions and take into account the uncertainties in
the couplings of the axial-vector mesons.
\vspace{-0.3cm}
}
\end{center}
\end{figure*}

\begin{figure*}[t]
\begin{center}
\includegraphics[clip=true,width=2\columnwidth,angle=0.]
{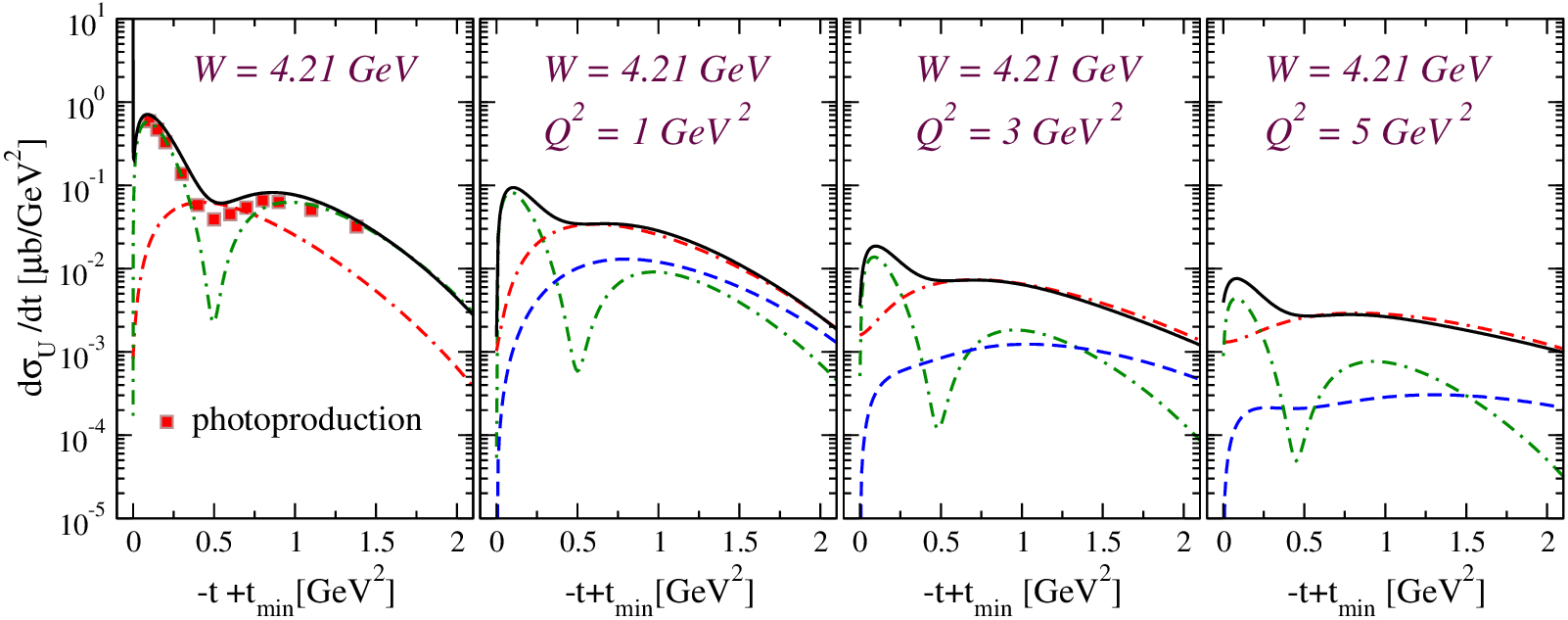}
\caption{\label{Pi0Hermes}
(Color online) $-t+t_{min}$ dependence of the differential cross section
  $d{\sigma}_{\rm T}/dt$ in photoproduction~\cite{Anderson:1971xh} ($Q^2=0$) and $d\sigma_{\rm U}/dt$ in
  electroproduction for different values of $Q^2=1$, 3 and 5~GeV in the
  kinematics of the HERMES experiment~\cite{Vandenbroucke:2007tc}. The solid
  curves are the model results and include the Regge and $s(u)$-channel 
 resonance contributions. The dash-dotted curves describe the Regge exchange
  and the dash-dash-dotted curves describe the resonance components,
  respectively. The dashed curves describe  the contribution of the longitudinal
  cross section $\varepsilon d\sigma_{\rm L}/dt$. 
\vspace{-0.7cm}
}
\end{center}
\end{figure*}

In general, a nonzero $\sigma_{\rm LT'}$ or the corresponding 
beam SSA $A_{\rm LU}(\phi)$, Eq.~(\ref{BSSA}), demands 
interference between single helicity flip and nonflip or double helicity flip 
amplitudes. 
Furthermore, $\sigma_{\rm LT'}$ is
proportional to the imaginary part of an interference between the \textsc{l/t}
photons and therefore sensitive to the relative phases of amplitudes.
In Regge models the asymmetry may result from 
Regge cut corrections to single Reggeon exchange~\cite{Ahmad:2008hp}. 
This way the amplitudes in the product acquire different phases and therefore 
relative imaginary parts. A nonzero beam SSA  can be also generated by the 
interference pattern of amplitudes where particles with opposite
parities are exchanged.
 In the following we discuss briefly the generic features of the beam SSA in the present
model. For the recent calculations of the beam SSA at JLAB in the
partonic and Regge models see Refs.~\cite{Ahmad:2008hp,Laget:2010za}.

In Fig.~\ref{BeamSpinAsymmPi0} 
we present our results for the azimuthal moment $A^{\sin(\phi)}_{\rm LU}$
associated with the beam SSA, Eq.~(\ref{BeamSSAmoment}), in the reaction 
$p(\vec{e},e'\pi^0)p$. The  CLAS/JLAB  data are from~\cite{DeMasi:2007id}.
Consider $A^{\sin(\phi)}_{\rm LU}$ generated by the exchange of 
Regge trajectories. In Fig.~\ref{BeamSpinAsymmPi0} the dashed curves describe  
the model results with $\omega$- and $\rho$- Regge trajectories alone. 
This simple Regge model results in a zero
$A^{\sin(\phi)}_{\rm LU}$ and therefore a zero beam SSA. 
Note that, in the charged pion production the addition of the unnatural
parity $a_1(1260)$-exchange generates by the interference with the natural parity 
$\rho(770)$ exchange a sizable $A^{\sin(\phi)}_{\rm LU}$ in both $\pi^+$ and $\pi^-$
channels. However, in the $\pi^0$ production the amplitude describing the
unnatural parity axial-vector $b_1$ and $h_1$  exchanges does not interfere with
the natural parity $\omega(\rho)$ exchanges. 
Therefore, there is no way to generate the nonzero beam SSA
in the model based only on the exchange of single Regge trajectories.

The $s(u)$-channel resonance contributions strongly influence the asymmetry parameter
$A^{\sin(\phi)}_{\rm LU}$. The solid curves describe the effect of nucleon
resonances only and are in good
agreement with data. The shaded bands describe the model results
which include the Regge and $s(u)$-channel resonance contributions. The band
takes into account the vector and axial-vector mesons. 
The width of the band reflects our estimations, see Fig.~\ref{Pi0photoAsymmetry}, of
couplings and form factors of the axial-vector mesons.
It indicates that in the present model the data do not require the presence of
large axial-vector exchanges.

\section{Deeply virtual $p(e,e'\pi^0)p$ at HERMES}
The HERMES collaboration at DESY already attempted 
to measure the exclusive $\pi^0$ electroproduction of protons in the DIS
region~\cite{Vandenbroucke:2007tc}. 
The $\pi^+$ data reported in Ref.~\cite{:2007an} have been analyzed using 
the present model in Ref.~\cite{KM}. In Fig.~\ref{Pi0Hermes} we present 
our results for the exclusive reaction $p(e,e'\pi^0)p$ in the kinematics at HERMES. 

We choose the value of $W=4.21$~GeV where the photoproduction
data exist~\cite{Anderson:1971xh}. Then the behavior of the electroproduction 
cross section  $d\sigma_{\rm
  U}/dt$ for the beam energy $E_e=27.7$~GeV and as a
function of $Q^2$ is shown for three bins $Q^2=1$, 3 and 5~GeV$^2$. These
results correspond to the
VMD cut-offs in the ${\gamma\omega\pi^0}$ and ${\gamma\rho\pi^0}$ form factors. 
As one can see, the very different $Q^2$ behavior of the Regge-exchange
(dash-dotted curves) and the resonance
contributions (dash-dash-dotted curves) washes out the dip region. 
Interestingly, contrary to the JLAB
data discussed above the forward region in the $\pi^0$ electroproduction at HERMES is dominated by 
the Regge-exchange contributions. Also the dashed curves
describe the contribution of longitudinal cross section 
$\varepsilon d\sigma_{\rm L}/dt$ to the total unseparated cross section 
$d\sigma_{\rm U}/dt$. Again, contrary to pQCD based approaches, in the present
model the longitudinal cross section
is marginally small and the cross section  at HERMES is
dominated by the conversion of the tranverse photons in $\sigma_{\rm T}$.

\section{Summary}
In summary, we have extended the model of Ref.~\cite{KM} and considered exclusive
electroproduction of $\pi^0$ off protons. The reaction amplitude
has been described by exchanges of vector $\omega(782)$, $\rho(770)$ and
axial-vector $h_1(1170)$  and $b_1(1235)$ Regge trajectories. The residual 
effect of $s$- and $u$-channel nucleon resonances has been taken into account using a dual
connection between the exclusive form factors and inclusive deep inelastic 
structure functions.
We have shown
that with these components and using the same model parameters as in the charged
pion production one can quantitatively describe the $\pi^0$ cross sections  and
beam SSA measured at JLAB. 

We have briefly discussed the real photon point  where the exchanges of Regge trajectories dominate.
Here the contribution of the $\omega(782)$-Regge trajectory plays an important
role and should be considered as an indispensable part of any model.  
As a novel feature we demonstrated an interesting effect of $s$- and
$u$-channel contributions around the region where the dip occurs.

However, in high $Q^2$ electroproduction the importance of Regge and nucleon
resonance reaction mechanisms is just opposite to that seen in
photoproduction. With increasing value of $Q^2$ the contributions of Regge exchanges
decrease strongly and the reaction cross section is dominated by the
excitation of nucleon resonances and corresponding large transverse
contributions. The model describes the measured cross
sections fairly well and clearly demonstrate the importance of 
resonances in the description of high $Q^2$ electroproduction data at JLAB. Similar 
effects of the nucleon resonances have been already discussed in the electroproduction of
charged pions.

We find that a model with $\omega$, $\rho$, $h_1$ and $b_1$
Regge exchanges does not generate any beam SSA. Since the 
axial-vector and vector meson exchanges do not interfere there is no way to generate the
beam SSA just based on exchanges of single Regge
trajectories. The positive beam SSA
observed in the experiment is a result of the resonance contributions. This is
again very similar to the charged pion production~\cite{KM} where the nucleon resonances 
are at the origin of the nonzero beam SSA observed in the experiment.

\begin{acknowledgments}
I am grateful to Prof. Ulrich Mosel for reading the manuscript and many useful comments.
This work was supported by DFG through TR16 and by BMBF.
\end{acknowledgments}

\end{document}